\documentclass[twocolumn,superscriptaddress,showpacs,preprintnumbers,amsmath,amssymb,aps,pre]{revtex4}
\usepackage{graphicx}
\usepackage{dcolumn}
\usepackage{bm}
\usepackage{fancyheadings}
\begin{document}
\title{Additional information for the paper `Attempt to distinguish long range temporal correlations from the statistics of the increments by natural time analysis' after its initial submission on February 25, 2006. Part II, Updated}
\author{P. A. Varotsos}
\email{pvaro@otenet.gr}
\affiliation{Solid State Section, Physics Department, University of Athens, Panepistimiopolis, Zografos 157 84,
Athens, Greece}
\affiliation{Solid Earth Physics Institute, Physics Department, University of Athens, Panepistimiopolis, Zografos 157 84, Athens, Greece}
\author{N. V. Sarlis}
\affiliation{Solid State Section, Physics Department, University of Athens, Panepistimiopolis, Zografos 157 84,
Athens, Greece}
\author{E. S. Skordas}
\affiliation{Solid State Section, Physics Department, University of Athens, Panepistimiopolis, Zografos 157 84,
Athens, Greece}
\affiliation{Solid Earth Physics Institute, Physics Department, University of Athens, Panepistimiopolis, Zografos 157 84, Athens, Greece}
\author{H. K. Tanaka}
\affiliation{Earthquake Prediction Research Center, Tokai University 3-20-1, Shimizu-Orido, Shizuoka 424-8610, Japan}
\author{M. S. Lazaridou}
\affiliation{Solid State Section, Physics Department, University of Athens, Panepistimiopolis, Zografos 157 84,
Athens, Greece}

\begin{abstract}
As mentioned in the preceding additional information (hereafter called Part I), a series of strong earthquakes with magnitudes between 5.2 and 5.9-units occurred during the two weeks period: 3 to 19 April, 2006 with epicenters lying at distances 80 to 100 km west of PAT station. Here, we show that the analysis in the natural time of the seismicity that occurred after the Seismic Electric Signals (SES) activity on February 13, 2006, specifies the occurrence time of the initiation of the aforementioned earthquake activity within a narrow range around two days. Furthermore, we provide the most recent information on some points mentioned in the main text.
\end{abstract}
\pacs{05.40.-a, 91.30.Dk, 05.45.Tp, 89.75.-k}
 \maketitle

According to the Athens observatory (the data of which will be used here),  a series of strong earthquakes (EQs) with magnitudes ranging from 5.2 to 5.9-units  occurred during the period: 3 to 19 April, 2006 , see Table I of Part I\cite{EPAPS1}. All their epicenters lie at distances of around 80 to 100km west of PAT station on the dipoles of which the SES activities depicted in Figs. 1(a) and 2(a) of the main text were recorded. 

\begin{figure}
\includegraphics{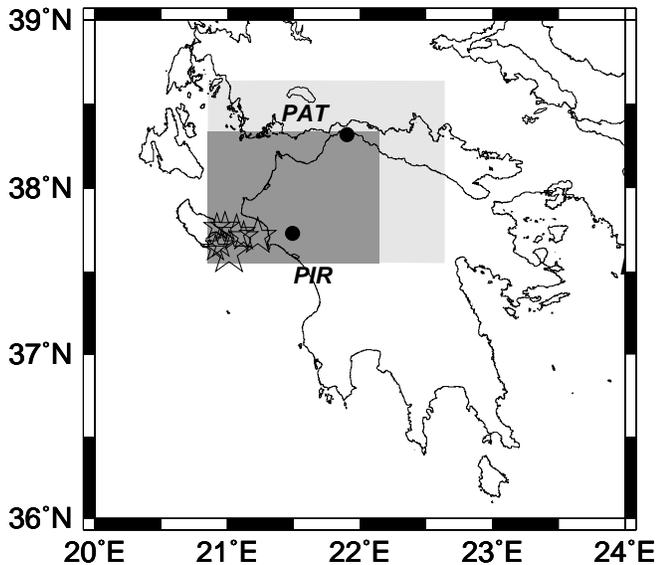}
\caption{Map of the area surrounding the  stations PAT and PIR (solid dots) and the epicenters of the strong EQs (stars) that occurred during the period: 3 to 19 April, 2006. The seismicity subsequent to the SES initiation on February 13, 2006, has been  studied in the gray shaded areas (the large and the small area are designated A and B, respectively).} \label{f1}
\end{figure}

\begin{figure*}
\includegraphics{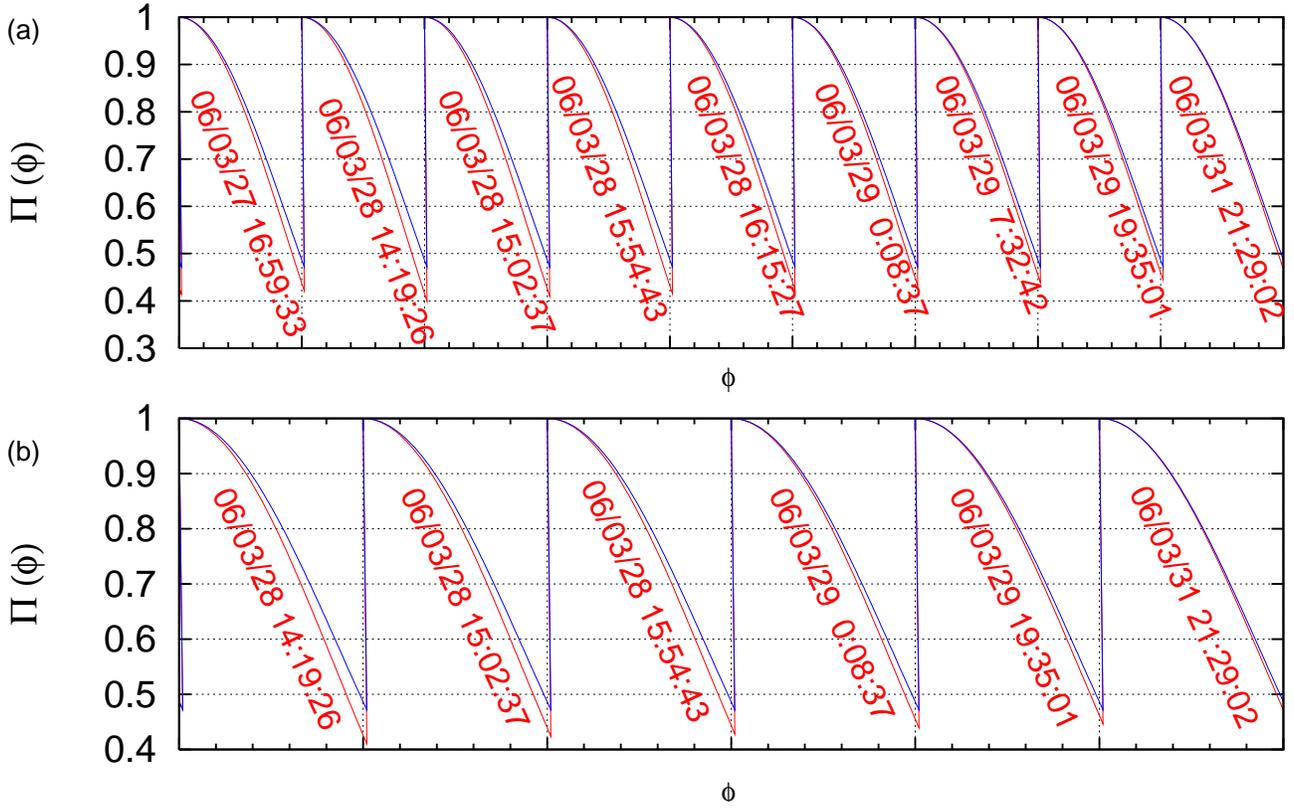}
\caption{(color)The normalized power spectrum(red) $\Pi (\phi )$  of the seismicity   as it evolves event by event (whose date and time of occurrence are written in each panel) after the 
initiation of the SES activity on February 13, 2006.  The two excerpts presented here refer to the period 27 to 31 March, 2006 and correspond to the area A $M_{thres}=3.0$ (a) and the area B $M_{thres}=2.8$ (b).  In each case only the spectrum in the area $\phi \in [0,0.5]$ (for the reasons 
discussed in Refs.\cite{NAT01,tan05}) is depicted (separated by the vertical dotted lines), whereas the  $\Pi (\phi )$ of Eq.(\ref{fasma}) is depicted by blue color. The minor horizontal ticks for $\phi$ are marked every 0.1.   }
\label{FigFasma}
\end{figure*}

\begin{figure}
\includegraphics{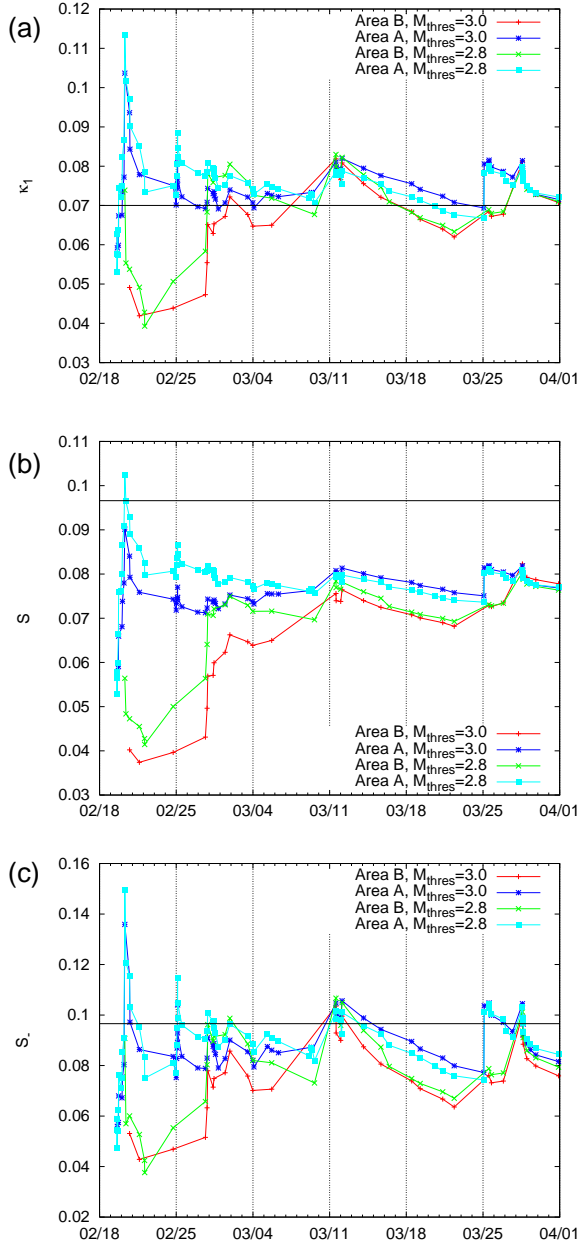}
\caption{(color)Evolution of the quantities  $\kappa_1$, $S$ and $S_{-}$. They are shown in (a), (b) and (c), respectively for two magnitude thresholds, i.e., $M \geq 2.8$ and $M \geq 3.0$,  for both areas A and B. After the event at 14:19 UT of March 28, the four curves (corresponding to the four combinations, i.e., resulting from the two areas and the two magnitude thresholds) almost collapse on the same curve. This points to the scale-invariance when approaching the critical point (see the text).}
\label{k1Sb}
\end{figure}

\begin{figure}
\includegraphics{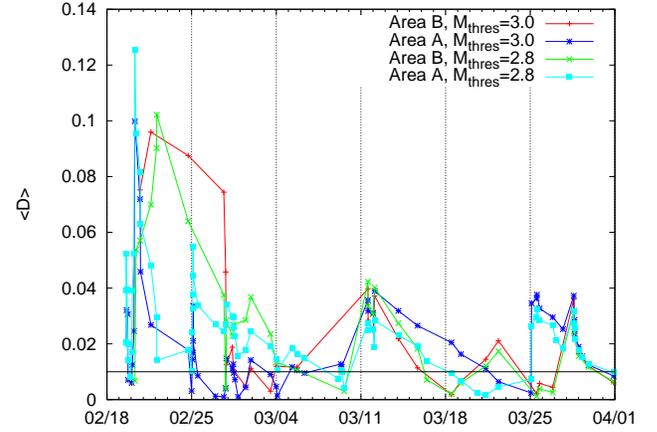}
\caption{(color)The average distance $\langle D \rangle$ between the calculated and the theoretical $\Pi(\phi )$ curves plotted here, for the sake of convenience, versus the conventional time. The calculation of $\langle D \rangle$ is made upon the occurrence of every consecutive earthquake when starting the calculation after the SES activity of February 13, 2006 (depicted in Fig.1(a) of the main text)
 for each of the two areas A and B by considering two magnitude thresholds 2.8 and 3.0.}
\label{Distance}
\end{figure}

\begin{figure*}
\includegraphics{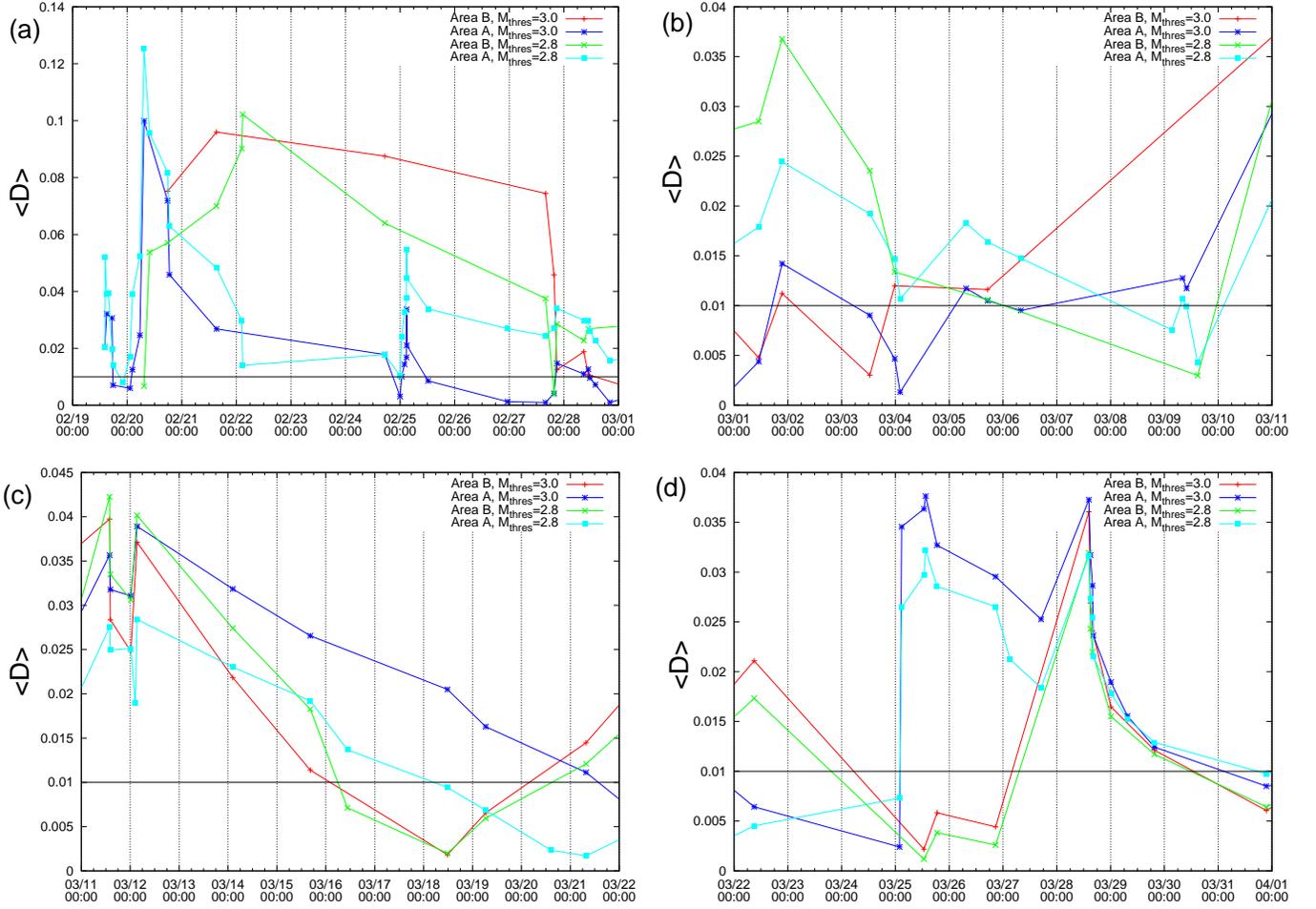}
\caption{(color) This figure presents, for the sake of clarity, four consecutive segments of Fig.\ref{Distance}. Note that in (d), after 14:19 UT of March 28 the four curves (corresponding to the four combinations resulting from the two areas and the two magnitude thresholds) almost collapse on the same curve. This points to the scale-invariance when approaching the critical point (see the text). }
\label{SuperSig}
\end{figure*}
\begin{figure}
\includegraphics{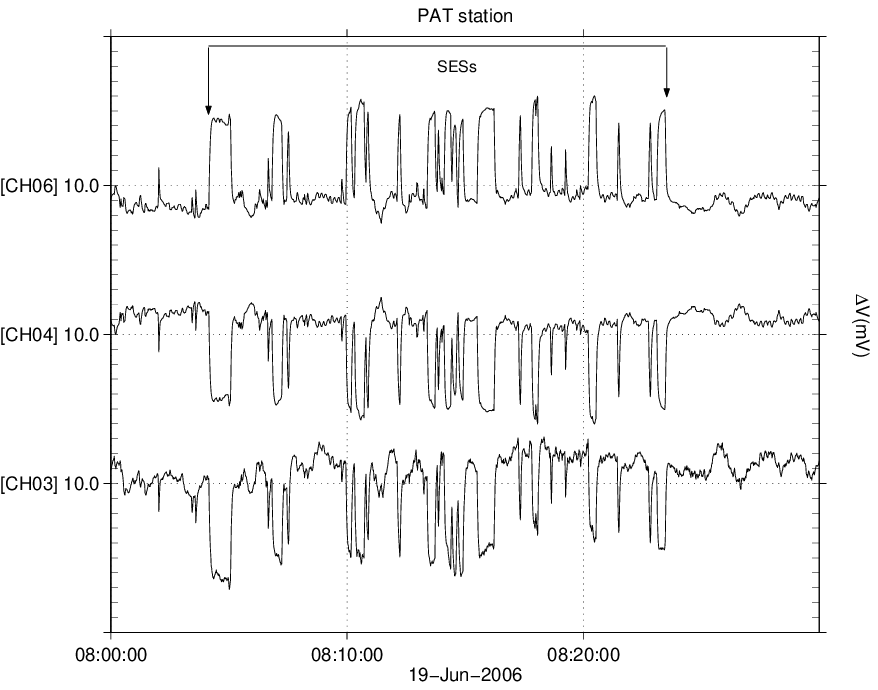}
\caption{The SES activity recently recorded at PAT station on June 19, 2006.}
\label{pat190606}
\end{figure}


We show that the occurrence time of the impending strong EQ activity can be estimated by following the procedure described in Refs.\cite{NAT01,tan05}. (We clarify that, during the last decade, preseismic information based on SES activities is published  {\em only} when the magnitude of the strongest EQ of the impending EQ activity is estimated to be -by means of the SES amplitude\cite{EPAPS}- comparable to 6.0 units or larger\cite{newbook}.)  We study how the seismicity evolved after the recording of the SES activity on February 13, 2006 (cf. This is the intense signals' activity that was classified as SES activity in the initially submitted version of the present paper on February 25, 2006), by considering either the area A:$N_{37.55}^{38.64}E_{20.85}^{22.64}$ or  the area B:$N_{37.55}^{38.34}E_{20.85}^{22.15}$, which 
 surround not only the EQ epicenters and the PAT station (see Fig.\ref{f1}) but also the PIR station at which a precursory GVEF started during the last week of February, 2006\cite{EPAPS1}. 
 If we set the natural time for seismicity zero at the initiation of the  SES activity on February 13, 2006, we form 
 time series of seismic events in natural time for various time windows as the number $N$ of consecutive (small) 
 EQs increases. We then compute the normalized power spectrum \cite{NAT01,tan05} in natural time $\Pi (\phi )$ for 
 each of the time windows. Excerpts of these results, which refer to the values deduced during the period 27 March to 1 April,2006, are depicted in red in Fig.\ref{FigFasma}.  In this figure,
 Fig.\ref{FigFasma}(a) corresponds to the
  area A  with magnitude threshold (herafter referring to the local magnitude ML or 
 the `duration' magnitude MD) $M_{thres}=3.0$, while Fig.\ref{FigFasma}(b) to the area B with $M_{thres}=2.8$. In the same figure, we plot in blue the power spectrum obeying 
 the relation\cite{NAT01,NAT02,NAT02A} 
\begin{equation}
\Pi ( \omega ) = \frac{18}{5 \omega^2}
-\frac{6 \cos \omega}{5 \omega^2}
-\frac{12 \sin \omega}{5 \omega^3}
\label{fasma}
\end{equation}
which holds when the system enters the {\em critical} stage ($\omega = 2\pi \phi$, where $\phi$ stands for the natural frequency\cite{NAT01,NAT02,NAT02A,newbook}). The date and the time of the occurrence of each small earthquake (with magnitude exceeding (or equal to) the aforementioned threshold) that occurred in each of the areas A and B, is also written in red in each panel (see also Table \ref{tab60}). 
An inspection of Fig.\ref{FigFasma} reveals that the red line approaches the blue line as $N$ increases and a {\em coincidence} occurs at the last small
event which had a magnitude 3.0 and occurred at 21:29 UT on March 31, 2006, i.e., roughly two days before the first strong EQ (00:50 UT on April 3,2006). To ensure that this coincidence is a {\em true} one\cite{NAT01,tan05,NAT02A,newbook} (see also below) we also calculate the evolution of the quantities $\kappa_1$,$S$ and $S_{-}$   and the results are depicted in Fig.\ref{k1Sb} for both magnitude thresholds 2.8 and 3.0 for each of the areas A and B.

The conditions for a coincidence to be considered as {\em true} are the following (e.g., see Ref.\cite{NAT01}, see also \cite{tan05,NAT02A,newbook}): First, the `average' distance $\langle D \rangle$ between the empirical and the theoretical $\Pi(\phi )$(i.e., the red and the blue line, respectively, in Fig.2) should be\cite{NAT01,tan05,NAT02,newbook} smaller than $10^{-2}$. See Fig.\ref{Distance} where we plot $\langle D \rangle$ versus the conventional time for the aforementioned two areas and the two magnitude thresholds (hence four combinations were studied in total). In order to better visualize the details of this figure, its four consecutive segments are enlarged and separately depicted in Fig.\ref{SuperSig}(a) to (d). Second, in the examples observed to date\cite{NAT01,tan05,NAT02A,newbook,EPAPS}, a few events {\em before} the coincidence leading to the strong EQ, the evolving $\Pi(\phi )$ has been found to approach that of Eq.(1), i.e., the blue one in Fig.2 , from {\em below} (cf. this reflects that during this approach the $\kappa_1$-value decreases as the number of events increases). 
In addition, both values $S$ and $S_{-}$ should be smaller than $S_u$ at the coincidence. Finally, 
since the process concerned is self-similar ({\em critical} dynamics), the time of the occurrence of the (true) coincidence  should {\em not} change, in principle, upon changing either the (surrounding)
area or the magnitude threshold used in the calculation. Note that in Fig.\ref{Distance} or Fig.\ref{SuperSig}(d), upon the occurrence of the aforementioned 
last small event of March 31, 2006, in both areas A and B and both magnitude thresholds(i.e., $M_{thres}=$2.8 and 3.0) their $\langle D \rangle$ values become smaller than $10^{-2}$.  Hence, this coincidence can be considered as {\em true}, while  other coincidences that occurred earlier (i.e., before March 31, 2006) have been found {\em not to be true ones} since they violate one or more of the aforementioned conditions(cf. typical examples of coincidences that are not true can be also found in Ref.\cite{EPAPS}).

The aforementioned strong earthquakes, the epicenters of which are shown in Fig.\ref{f1}, lasted from 3 to 19 April, 2006. They were preceded by the SES activity of February 13, 2006, as well as (some of them by) the one recorded on April 13, 2006 that were presented in the main text. We now comment on what happened after this seismic activity. First, two additional SES activities have been recorded at PAT on April 19 and 21, 2006, see Figs. 2(b) and (c) of Part I\cite{EPAPS1}. They have been followed by two EQs that occurred at seismic areas {\em different} than the previously active one (as expected\cite{EPAPS1}): (a) At 06:16 UT on May 5, 2006, an EQ with magnitude 4.8 and an epicenter at $38.28^oN 22.63^oE$, i.e., almost 60 km E of PAT station and (b) at 23:14 UT of May 25, 2006, an EQ with magnitude 5.3 with an epicenter at  $36.87^oN 20.34^oE$, i.e., almost 100 km SW of PIR station. Second, an intense signal (see Fig.\ref{pat190606}) has been recently recorded at PAT on June 19, 2006, which is similar to the one depicted in Fig.1(a) of the main text. It has been also classified as SES activity by following a procedure similar to that explained in the main text. The analysis of the evolving seismicity subsequent to the latter SES activity, in order to specify the occurrence time of the impending strong earthquake (with an epicenter possibly lying within the area A,but see also the next paragraph) -in a fashion similar to that explained here- is in progress. Third, the following phenomenon was noticed 
at the station KER (i.e., Keratea station located close to Athens) after the submission of the revised version of the present paper on June 30,2006. Figure 7 depicts the data of three, for example, electric dipoles during the period from 10 May to 18 July, 2006. A gradual deviation from the background level presumably develops during the last few weeks, which suggests the evolution of a precursory GVEF (cf. in order for this variation to be finally classified as GVEF, however, the channels should return at their previous levels, a fact which is still unknown). Recall that, as already mentioned in Part I, such a phenomenon is observed only for EQs with magnitude larger than 5.5 when the (future) epicenter happens to lie at a small distance (some tens of km) from the measuring station. Is this GVEF at KER interconnected to the detection of the aforementioned SES activity at PAT on June 19, 2006? (The other alternative, of course, is that these two phenomena correspond to separate events.) To investigate such a possibility the analysis of the evolving seismicity subsequent  to the latter SES activity -in order to specify the occurrence time of the impending strong EQ by means of the method explained above-  is also currently carried out by considering an area extending from PAT to KER (besides the one within the area A). The distinction between the two
alternatives will be achieved, in the course of time, upon checking which of the two current analyses will exhibit a scale 
invariant feature (similar to the one visualized in Figs.4 and 5(d)) when approaching the critical point.

Let us briefly summarize: First, the occurrence time of the initiation of the strong seismic activity, that lasted from 3 to 19 April, 2006 at an epicentral region 80 to 100 km west of PAT, can be specified within a narrow range around 2 days. This is so, because the power spectrum in natural time of the evolving seismicity after the SES activity of February 13, 2006, collapses on the one expected for critical dynamics at 21:29 UT on March 31, 2006, i.e., almost two days before the occurrence time of the 5.3 earthquake of April 3, 2006. Second, the two SES activities that were observed {\em after} the aforementioned strong seismic activity, have been also followed by two earthquakes on May 5 and 25, 2006, the epicenters of which lie clearly outside the previously active region, as expected\cite{EPAPS1}. Third, the most recent electric data are presented.

\begin{figure}
\includegraphics{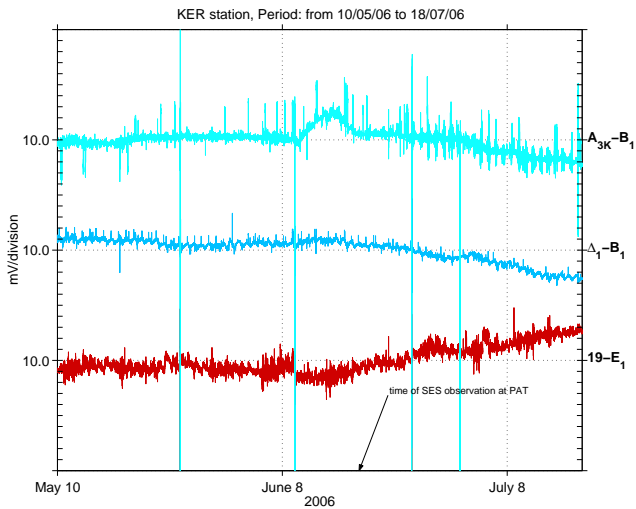}
\caption{(Color) Plot of more than two months data recorded at three electric dipoles operating at KER (cf. the sites of these
electrodes are shown in Fig.1.4.2 of Ref.\cite{newbook}). Period:May 10, 2006 to July 18, 2006 (see the text). For the reader's convenience, the occurrence time of the SES activity at PAT on June 19, 2006 is also marked by an arrow.}
\end{figure} 
\pagebreak
\begin{table*}
\caption{The catalogue ($M \geq 2.8$) of the Institute of Geodynamics of the National Observatory of Athens for the area A under discussion during the following period: From the initiation of the SES activity on February 13, 2006 (Fig.1 of the main text) until the occurrence of the Ms(ATH)=5.3 earthquake on April 3, 2006. Note that Ms(ATH)=M+0.5, where M stands for ML or MD.} \label{tab60}
\begin{ruledtabular}
\begin{tabular}{ccccccccc}
No & Year & Month & Day & UT  &
Lat.($^o$N) & Lon.($^o$E) & depth(km) & M  \\
\hline
   1 & 2006 &  2 &  15 &  6:37:45 &  37.99 &  22.02 &  13.0 &   3.0 \\
   2 & 2006 &  2 &  18 & 17:23:15 &  38.37 &  21.97 &  32.0 &   3.1 \\
   3 & 2006 &  2 &  19 &  1:16:49 &  38.43 &  21.77 &  29.0 &   3.1 \\
   4 & 2006 &  2 &  19 & 13:32:10 &  38.41 &  21.98 &   5.0 &   3.2 \\
   5 & 2006 &  2 &  19 & 13:55:54 &  38.35 &  22.02 &  20.0 &   3.3 \\
   6 & 2006 &  2 &  19 & 14:08:27 &  38.22 &  21.77 &  34.0 &   3.0 \\
   7 & 2006 &  2 &  19 & 14:12:59 &  38.36 &  21.93 &  45.0 &   2.8 \\
   8 & 2006 &  2 &  19 & 14:56:39 &  38.33 &  22.03 &  32.0 &   3.0 \\
   9 & 2006 &  2 &  19 & 15:42:32 &  38.32 &  22.15 &  31.0 &   2.9 \\
  10 & 2006 &  2 &  19 & 17:19:22 &  38.29 &  22.25 &  31.0 &   3.0 \\
  11 & 2006 &  2 &  19 & 17:44:57 &  38.36 &  22.01 &  26.0 &   3.1 \\
  12 & 2006 &  2 &  19 & 22:09:41 &  38.38 &  21.98 &  24.0 &   2.9 \\
  13 & 2006 &  2 &  20 &  1:13:47 &  38.41 &  21.75 &  23.0 &   3.0 \\
  14 & 2006 &  2 &  20 &  2:16:47 &  38.34 &  22.03 &  17.0 &   3.1 \\
  15 & 2006 &  2 &  20 &  5:34:58 &  38.41 &  21.92 &  28.0 &   3.1 \\
  16 & 2006 &  2 &  20 &  7:21:60 &  37.85 &  21.15 &  14.0 &   3.6 \\
  17 & 2006 &  2 &  20 &  9:47:41 &  38.31 &  21.99 &  32.0 &   2.9 \\
  18 & 2006 &  2 &  20 & 17:45:14 &  38.33 &  21.97 &  18.0 &   3.3 \\
  19 & 2006 &  2 &  20 & 18:33:49 &  38.36 &  21.96 &  32.0 &   3.1 \\
  20 & 2006 &  2 &  21 & 15:15:48 &  38.23 &  21.59 &  39.0 &   3.1 \\
  21 & 2006 &  2 &  22 &  2:25:19 &  38.34 &  21.99 &  26.0 &   2.8 \\
  22 & 2006 &  2 &  22 &  2:46:46 &  38.34 &  22.00 &  22.0 &   2.8 \\
  23 & 2006 &  2 &  24 & 17:16:32 &  37.92 &  21.15 &  12.0 &   3.2 \\
  24 & 2006 &  2 &  25 &  0:01:08 &  38.57 &  21.65 &  13.0 &   3.0 \\
  25 & 2006 &  2 &  25 &  0:44:23 &  38.61 &  21.52 &  22.0 &   3.3 \\
  26 & 2006 &  2 &  25 &  1:55:25 &  38.54 &  21.79 &  15.0 &   3.3 \\
  27 & 2006 &  2 &  25 &  2:47:35 &  38.53 &  21.83 &  18.0 &   3.3 \\
  28 & 2006 &  2 &  25 &  2:54:01 &  38.53 &  21.82 &  20.0 &   3.5 \\
  29 & 2006 &  2 &  25 &  2:58:49 &  38.42 &  22.20 &  10.0 &   3.1 \\
  30 & 2006 &  2 &  25 & 12:18:39 &  38.52 &  22.16 &   4.0 &   3.0 \\
  31 & 2006 &  2 &  26 & 23:14:15 &  38.40 &  22.01 &   4.0 &   3.1 \\
  32 & 2006 &  2 &  27 & 15:58:47 &  37.99 &  21.94 &  27.0 &   3.2 \\
  33 & 2006 &  2 &  27 & 19:41:00 &  37.98 &  21.99 &  22.0 &   3.3 \\
  34 & 2006 &  2 &  27 & 21:00:59 &  38.05 &  21.93 &  28.0 &   3.4 \\
  35 & 2006 &  2 &  28 &  8:48:46 &  38.01 &  22.08 &  20.0 &   3.2 \\
  36 & 2006 &  2 &  28 & 10:46:37 &  38.01 &  21.99 &  18.0 &   3.3 \\
  37 & 2006 &  2 &  28 & 11:27:33 &  38.00 &  22.28 &  39.0 &   3.2 \\
  38 & 2006 &  2 &  28 & 13:52:53 &  38.43 &  21.15 &  38.0 &   3.2 \\
  39 & 2006 &  2 &  28 & 20:13:51 &  38.00 &  22.22 &  35.0 &   3.0 \\
  40 & 2006 &  3 &   1 & 11:00:48 &  37.98 &  21.99 &  18.0 &   3.3 \\
  41 & 2006 &  3 &   1 & 21:27:40 &  38.02 &  22.02 &   7.0 &   3.4 \\
  42 & 2006 &  3 &   3 & 12:34:51 &  37.79 &  21.15 &  26.0 &   3.1 \\
  43 & 2006 &  3 &   3 & 23:41:13 &  37.94 &  21.89 &  56.0 &   3.1 \\
  44 & 2006 &  3 &   4 &  2:06:53 &  38.44 &  21.80 &  27.0 &   3.1 \\
  45 & 2006 &  3 &   5 &  7:35:24 &  38.40 &  21.99 &  17.0 &   3.4 \\
  46 & 2006 &  3 &   5 & 17:10:43 &  38.34 &  22.02 &  27.0 &   3.2 \\
  47 & 2006 &  3 &   6 &  7:57:06 &  38.41 &  21.96 &  31.0 &   3.2 \\
  48 & 2006 &  3 &   9 &  3:16:08 &  38.62 &  22.38 &  21.0 &   2.8 \\
  49 & 2006 &  3 &   9 &  8:10:05 &  37.58 &  22.59 &  68.0 &   3.3 \\
  50 & 2006 &  3 &   9 &  9:56:47 &  38.54 &  21.14 &  35.0 &   3.2 \\
  51 & 2006 &  3 &   9 & 14:51:03 &  38.20 &  21.96 &  21.0 &   2.9 \\
  52 & 2006 &  3 &  11 & 13:51:31 &  37.86 &  21.01 &   5.0 &   3.6 \\
  53 & 2006 &  3 &  11 & 14:15:15 &  37.86 &  21.01 &  32.0 &   3.2 \\
  54 & 2006 &  3 &  12 &  0:17:22 &  37.80 &  20.98 &   5.0 &   3.3 \\
  55 & 2006 &  3 &  12 &  2:26:50 &  38.13 &  22.63 &  13.0 &   2.9 \\
      \end{tabular}
\end{ruledtabular}
\end{table*}

\addtocounter{table}{-1}
\begin{table*}
\caption{Continued}
\begin{ruledtabular}

\begin{tabular}{ccccccccc}
No & Year & Month & Day & UT  &
Lat.($^o$N) & Lon.($^o$E) & depth(km) & M  \\
\hline  
   56 & 2006 &  3 &  12 &  3:22:22 &  37.88 &  20.99 &  27.0 &   3.5 \\
  57 & 2006 &  3 &  14 &  2:20:00 &  38.33 &  21.89 &  30.0 &   3.0 \\
  58 & 2006 &  3 &  15 & 16:22:00 &  38.33 &  22.05 &  35.0 &   3.1 \\
  59 & 2006 &  3 &  16 & 10:34:05 &  38.26 &  22.10 &  35.0 &   2.9 \\
  60 & 2006 &  3 &  18 & 11:40:57 &  38.26 &  21.50 &  34.0 &   3.0 \\ 
  61 & 2006 &  3 &  19 &  6:26:46 &  37.73 &  21.21 &  12.0 &   3.1 \\
  62 & 2006 &  3 &  20 & 14:30:39 &  38.56 &  21.51 &  22.0 &   2.9 \\
  63 & 2006 &  3 &  21 &  7:43:18 &  38.12 &  21.85 &  13.0 &   3.0 \\
  64 & 2006 &  3 &  22 &  8:57:56 &  37.64 &  21.51 &  31.0 &   3.0 \\
  65 & 2006 &  3 &  25 &  1:49:37 &  38.09 &  22.64 &  13.0 &   3.0 \\
  66 & 2006 &  3 &  25 &  2:46:38 &  38.64 &  21.80 &  23.0 &   3.7 \\
  67 & 2006 &  3 &  25 & 12:42:23 &  38.34 &  21.97 &  24.0 &   3.4 \\
  68 & 2006 &  3 &  25 & 13:29:59 &  38.61 &  21.81 &  23.0 &   3.4 \\
  69 & 2006 &  3 &  25 & 18:34:12 &  37.83 &  21.18 &  15.0 &   3.1 \\
  70 & 2006 &  3 &  26 & 20:43:11 &  37.87 &  21.06 &  25.0 &   3.2 \\
  71 & 2006 &  3 &  27 &  3:06:55 &  38.45 &  21.88 &  28.0 &   2.8 \\
  72 & 2006 &  3 &  27 & 16:59:33 &  38.63 &  21.55 &   5.0 &   3.1 \\
  73 & 2006 &  3 &  28 & 14:19:26 &  37.63 &  21.42 &  25.0 &   3.6 \\
  74 & 2006 &  3 &  28 & 15:02:37 &  37.88 &  21.08 &  10.0 &   3.0 \\
  75 & 2006 &  3 &  28 & 15:54:43 &  38.32 &  21.99 &  14.0 &   3.2 \\
  76 & 2006 &  3 &  28 & 16:15:27 &  38.36 &  21.93 &  28.0 &   3.0 \\
  77 & 2006 &  3 &  29 &  0:08:37 &  38.33 &  22.00 &   9.0 &   3.0 \\
  78 & 2006 &  3 &  29 &  7:32:42 &  38.37 &  22.00 &  27.0 &   3.1 \\
  79 & 2006 &  3 &  29 & 19:35:01 &  38.32 &  21.98 &  29.0 &   3.1 \\
  80 & 2006 &  3 &  31 & 21:29:02 &  38.31 &  22.12 &  29.0 &   3.0 \\
  81 & 2006 &  4 &   2 & 16:53:23 &  38.32 &  22.09 &  37.0 &   3.3 \\
  82 & 2006 &  4 &   2 & 21:29:50 &  38.61 &  21.87 &  38.0 &   3.0 \\
  {\bf 83} & {\bf 2006} & {\bf 4 }&  {\bf 3} &  {\bf 0:49:42} &  {\bf 37.59} &  {\bf 20.95} &  {\bf 20.0} &  {\bf 4.8 }  
\end{tabular}
\end{ruledtabular}
\end{table*}


\begin{thebibliography}{7}
\expandafter\ifx\csname natexlab\endcsname\relax\def\natexlab#1{#1}\fi
\expandafter\ifx\csname bibnamefont\endcsname\relax
  \def\bibnamefont#1{#1}\fi
\expandafter\ifx\csname bibfnamefont\endcsname\relax
  \def\bibfnamefont#1{#1}\fi
\expandafter\ifx\csname citenamefont\endcsname\relax
  \def\citenamefont#1{#1}\fi
\expandafter\ifx\csname url\endcsname\relax
  \def\url#1{\texttt{#1}}\fi
\expandafter\ifx\csname urlprefix\endcsname\relax\def\urlprefix{URL }\fi
\providecommand{\bibinfo}[2]{#2}
\providecommand{\eprint}[2][]{\url{#2}}

\bibitem[{EPA({\natexlab{a}})}]{EPAPS1}
\eprint{See the document {\tt Suplinfo.pdf} deposited in the same, as this
  file, EPAPS directory on April 25, 2006. This document may be retrieved via
  the EPAPS homepage (\url{http://www.aip.org/pubservs/epaps.html}) or from
  \url{ftp.aip.org} in the directory /epaps/. See the EPAPS homepage for more
  information.}

\bibitem[{\citenamefont{Varotsos et~al.}(2001)\citenamefont{Varotsos, Sarlis,
  and Skordas}}]{NAT01}
\bibinfo{author}{\bibfnamefont{P.~A.} \bibnamefont{Varotsos}},
  \bibinfo{author}{\bibfnamefont{N.~V.} \bibnamefont{Sarlis}},
  \bibnamefont{and} \bibinfo{author}{\bibfnamefont{E.~S.}
  \bibnamefont{Skordas}}, \bibinfo{journal}{Practica of Athens Academy}
  \textbf{\bibinfo{volume}{76}}, \bibinfo{pages}{294} (\bibinfo{year}{2001}).

\bibitem[{\citenamefont{Varotsos et~al.}(2005)\citenamefont{Varotsos, Sarlis,
  Tanaka, and Skordas}}]{tan05}
\bibinfo{author}{\bibfnamefont{P.}~\bibnamefont{Varotsos}},
  \bibinfo{author}{\bibfnamefont{N.}~\bibnamefont{Sarlis}},
  \bibinfo{author}{\bibfnamefont{H.}~\bibnamefont{Tanaka}}, \bibnamefont{and}
  \bibinfo{author}{\bibfnamefont{E.}~\bibnamefont{Skordas}},
  \bibinfo{journal}{Phys. Rev. E} \textbf{\bibinfo{volume}{72}},
  \bibinfo{pages}{041103} (\bibinfo{year}{2005}).

\bibitem[{EPA({\natexlab{b}})}]{EPAPS}
\eprint{See EPAPS Document No. E-PLEEE8-73-134603 for additional information.
  This document may be retrieved via the EPAPS homepage
  (\url{http://www.aip.org/pubservs/epaps.html}) or from \url{ftp.aip.org} in
  the directory /epaps/. See the EPAPS homepage for more information.}

\bibitem[{\citenamefont{Varotsos}(2005)}]{newbook}
\bibinfo{author}{\bibfnamefont{P.}~\bibnamefont{Varotsos}},
  \emph{\bibinfo{title}{The Physics of Seismic Electric Signals}}
  (\bibinfo{publisher}{TERRAPUB}, \bibinfo{address}{Tokyo},
  \bibinfo{year}{2005}).

\bibitem[{\citenamefont{Varotsos
  et~al.}(2002{\natexlab{a}})\citenamefont{Varotsos, Sarlis, and
  Skordas}}]{NAT02}
\bibinfo{author}{\bibfnamefont{P.~A.} \bibnamefont{Varotsos}},
  \bibinfo{author}{\bibfnamefont{N.~V.} \bibnamefont{Sarlis}},
  \bibnamefont{and} \bibinfo{author}{\bibfnamefont{E.~S.}
  \bibnamefont{Skordas}}, \bibinfo{journal}{Phys. Rev. E}
  \textbf{\bibinfo{volume}{66}}, \bibinfo{pages}{011902}
  (\bibinfo{year}{2002}{\natexlab{a}}).

\bibitem[{\citenamefont{Varotsos
  et~al.}(2002{\natexlab{b}})\citenamefont{Varotsos, Sarlis, and
  Skordas}}]{NAT02A}
\bibinfo{author}{\bibfnamefont{P.}~\bibnamefont{Varotsos}},
  \bibinfo{author}{\bibfnamefont{N.}~\bibnamefont{Sarlis}}, \bibnamefont{and}
  \bibinfo{author}{\bibfnamefont{E.}~\bibnamefont{Skordas}},
  \bibinfo{journal}{Acta Geophys. Pol.} \textbf{\bibinfo{volume}{50}},
  \bibinfo{pages}{337} (\bibinfo{year}{2002}{\natexlab{b}}).

\end{thebibliography}

\end{document}